\documentclass[twocolumn]{aastex62}
\usepackage[encapsulated]{CJK}
\usepackage{ucs}
\usepackage[utf8x]{inputenc}

\newcommand{\cntext}[1]{\begin{CJK}{UTF8}{gbsn}#1\end{CJK}}

\received{}
\revised{}
\accepted{\today}

\shorttitle{X-ray emission from NGC\,5189}
\shortauthors{Toal\'{a}, Montez \& Karovska}

\begin{document}

\title{\bf A carbon-rich hot bubble in the planetary nebula NGC\,5189}

\correspondingauthor{Jes\'{u}s A. Toal\'{a}}
\email{j.toala@irya.unam.mx}

\author[0000-0002-5406-0813]{Jes\'{u}s\,A.\,Toal\'{a}\cntext{(杜宇君)}}
\affiliation{Instituto de Radioastronomía y Astrofísica, UNAM Campus Morelia, Apartado Postal 3-72, Morelia 58090, Michoacán, Mexico}

\author[0000-0002-6752-2909]{Rodolfo\,Montez\,Jr}
\affil{Center for Astrophysics $\vert$ Harvard \& Smithsonian, 60 Garden Street, Cambridge, MA 02138, USA}

\author[0000-0003-1769-9201]{Margarita Karovska}
\affil{Center for Astrophysics $\vert$ Harvard \& Smithsonian, 60 Garden Street, Cambridge, MA 02138, USA}

\begin{abstract}

We present the discovery of extended X-ray emission from the planetary
nebula (PN) NGC\,5189 around the [WO1]-type WD\,1330-657 with {\it
  XMM-Newton}. The X-ray-emitting gas fills the cavities detected in
the {\it Hubble Space Telescope} [O\,{\sc iii}] narrow-band image and
presents a limb-brightened morphology towards the outer edges of the
east and west lobes. The bulk of the X-ray emission is detected in the
soft (0.3--0.7~keV) band with the {\it XMM-Newton} EPIC spectra
dominated by the C\,{\sc vi} Ly$\alpha$ line at 0.37~keV
(=33.7~\AA). Spectral analysis resulted in carbon and neon abundances
38 and 6 times their solar values, with a plasma temperature of
$kT=0.14\pm0.01$~keV ($T=1.6\times10^{6}$~K) and X-ray luminosity of
$L_\mathrm{X}=(2.8\pm0.8)\times10^{32}$~erg~s$^{-1}$. NGC\,5189 is an
evolved and extended PN ($\lesssim$0.70~pc in radius), thus, we
suggest that the origin of its X-ray emission is consistent with the
born-again scenario in which the central star becomes carbon-rich
through an eruptive {\it very late thermal pulse}, subsequently
developing a fast, carbon-rich wind powering the X-ray emission
as suggested for A\,30 and A\,78.

\end{abstract}



\section{Introduction}
\label{sec:intro}

Hot bubbles in planetary nebulae (PNe) are one of the direct
confirmations of the interacting stellar wind model
\citep[][]{Kreysing1992,Apparao1989,Guerrero2000}. The formation of
hot bubbles has been described as the direct interaction of the fast
wind from the central star (CSPN) with the previously ejected dense
and slow asymptotic giant branch (AGB) material. The fast wind
\citep[$v_{\infty}$=500--4000~km~s$^{-1}$;][]{Guerrero2013} slams into
the AGB material producing an adiabatically-shocked region around the
CSPN with temperatures in excess to 10$^{7}$~K
\citep[see][]{Dyson1997,Volk1985}.

High-quality X-ray observations, such as those obtained with {\it
  XMM-Newton} and {\it Chandra}, have unveiled in unprecedented detail
the distribution and physical properties of hot bubbles in PNe
\citep[see][and references
  therein]{Kastner2000,Guerrero2005,Gruendl2006,2008ApJ...672..957K,2013ApJ...767...35R}. Hot
bubbles appear to fill the inner cavities in PNe, and when the bubble
is resolved or the photon count rate is high, the X-ray-emitting gas
appears to be limb-brightened \citep[e.g.,][]{Chu2001,Montez2005}. The
estimated plasma temperatures derived from spectral fitting are a few
times 10$^{6}$~K, which is at least an order of magnitude below
theoretical expectations.

This temperature discrepancy has been known and widely discussed in
the past decades
\citep[e.g.,][]{2003ApJ...583..368S,Steffen2008,Toala2018}.  Two
possible physical mechanisms that might reduce the temperature of the
hot bubble have been proposed in the literature: i) thermal
conductivity and ii) hydrodynamical mixing. In the first case, hot
electrons transfer energy to the outer nebular material; the inclusion
of cold (10$^{4}$~K) material raises the density and lowers the
temperature at its outer edge
\citep{Weaver1977,Steffen2008}. Secondly, multi-dimensional numerical
simulations predict that the wind-wind interaction region will produce
clumps and filaments due to Rayleigh-Taylor and thin shell
instabilities naturally mixing the outer nebular material \citep[][and
  references therein]{Toala2018}.

The {\it Chandra} planetary nebula Survey ({\sc ChanPlaNS}) has been
designed to study the origins and properties of X-ray emission from
PNe in the solar neighborhood \citep[$d<2$~kpc;][]{Kastner2012}. Among
the main results of the {\sc ChanPlaNS} is that mostly young (age $<$
5000~yr) and compact (radius $<$ 0.2~pc) PNe will be detected in
X-rays. Furthermore, most of the targets that exhibit extended X-ray
emission are those PNe that drive powerful winds, namely those that
harbor [Wolf-Rayet]([WR])-type central stars
\citep{Kastner2012,Freeman2014}.  {\sc ChanPlaNS} has also peered into
the properties of CSPNe showing that there are two classes of CSPN
with hard X-ray emission \citep[see][]{Montez2015}. The first ones,
CSPN with high-temperature plasmas and high X-ray luminosities, are
correlated to active binary companions \citep[][]{Montez2010}. The
second class exhibit lower-plasma temperature with their X-ray and
bolometric luminosities following the
$L_\mathrm{X}/L_\mathrm{bol}\sim10^{-7}$ relation, similar to massive
hot stars \citep{Pallavicini1981,Nebot2018}, for which it has been
suggested that the X-ray emission is due to self-shocking winds.

\begin{figure}
\begin{center}
  \includegraphics[angle=0,width=\linewidth]{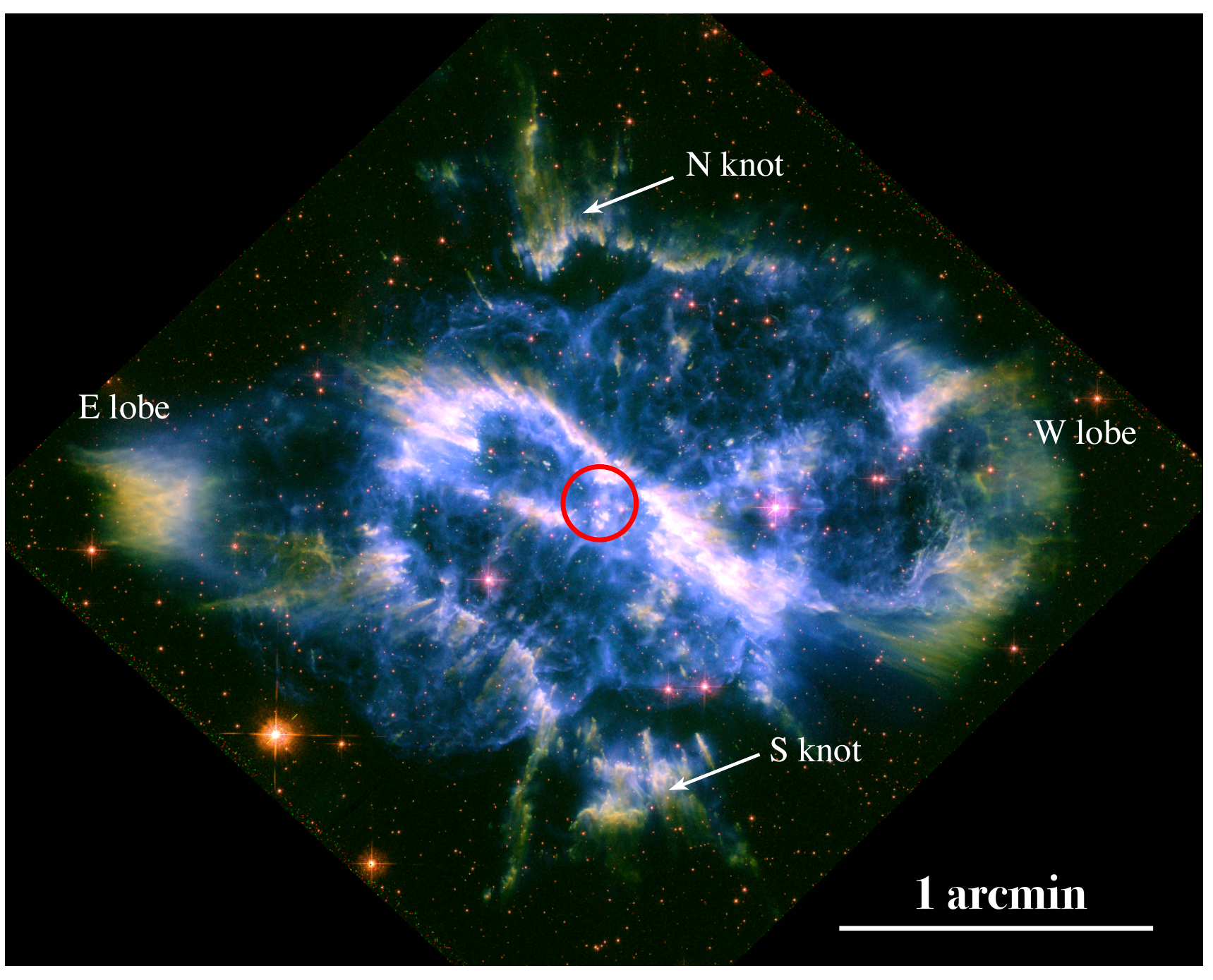}
\label{fig:NGC5189_opt}
\caption{Color-composite {\it HST} image of NGC\,5189. Red, green, and
  blue correspond to [S\,{\sc ii}], H$\alpha$, and [O\,{\sc iii}]
  emission, respectively. The position of the progenitor star is shown
  with a circular aperture. North is up, east to the left.}
\end{center}
\end{figure}

In this paper we present the discovery of a hot bubble within the PN
NGC\,5189 (a.k.a. PN\,G307.2$-$03.4) which harbors the [WO1]-type star
WD\,1330-657 \citep[][]{Acker2003}. This PNe has an intricate
morphology (see Fig.~\ref{fig:NGC5189_opt}) with at least three pairs
of bipolar lobes protruding from the central region that is mainly
traced by the [O\,{\sc iii}] emission plus a pair of low-ionization
structures located at the northern and southern regions (hereafter
knot N and S).  \citet{Sabin2012} reported the presence of an apparent
toroidal structure around the CSPN through near- and mid-IR images and
suggested that this was a dust-rich structure that could have
interacted with the CSPN fast wind to produce the expansion of the
lobes in NGC\,5189 along the E-W direction.  Recently,
\citet{Danehkar2018} showed the ionization structure of NGC\,5189
based on the analysis of {\it Hubble Space Telescope} ({\it HST}) WFC3
observations, exposing the presence of low-ionization structures
within this PN in great detail.

This paper is organized as follows. In Section~2 we present our {\it
  XMM-Newton} observations and data preparation. In Section~3 we
present our results. A discussion is presented in Section~4 and,
finally, the conclusions are presented in Section~5.

\section{{\it XMM-Newton} Observations}

NGC\,5189 was observed with {\it XMM-Newton} during 2018-02-05 with
the three European Photon Imaging Cameras (EPIC) pn, MOS1, and MOS2
(PI: R.\,Montez; Obs.\,ID.: 0801960101). The three EPIC cameras were
operated in the full frame mode with the thin optical blocking
filter. The total observing times for the EPIC pn, MOS1, and MOS2 were
84.95~ks, 86.56~ks, and 86.53~ks, respectively.

\begin{figure*}
\begin{center}
  \includegraphics[angle=0,width=\linewidth]{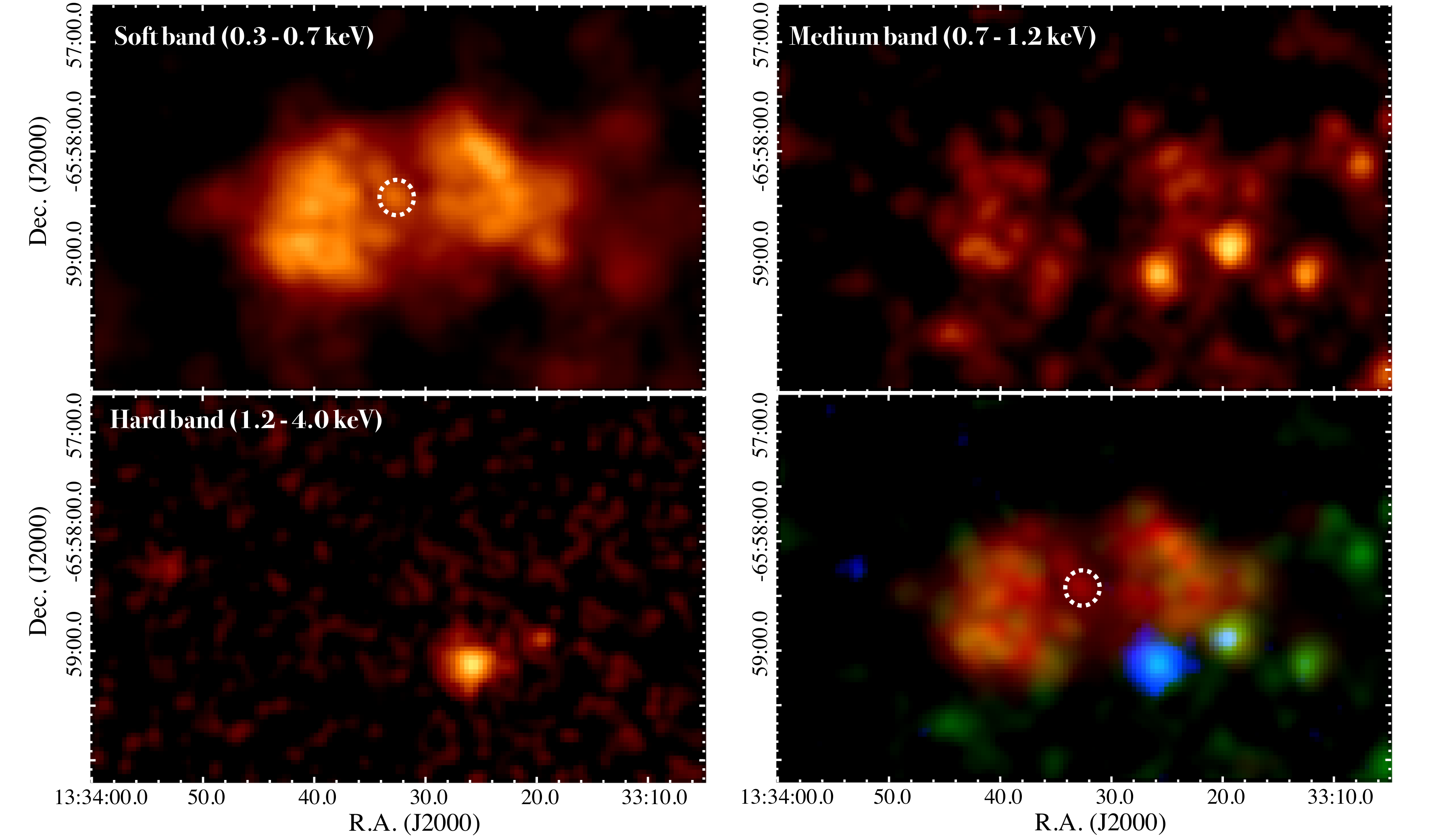}
\label{fig:NGC5189_XRAYS}
\caption{{\it XMM-Newton} EPIC (pn+MOS1+MOS2) images of NGC\,5189. The
  bottom right panel shows a color-composite image combining the soft
  (red), medium (green), and hard (blue) bands. The position of the
  CSPN is shown with a circular white dashed-line aperture in the
  upper left and lower right panels.}
\end{center}
\end{figure*}

The observations were processed with the {\it XMM-Newton} Science
Analysis System (SAS) version 17.0
\citep[][]{Gabriel2004}\footnote{The Users Guide to the {\it
    XMM-Newton} SAS can be found
  in:\\ \url{https://xmm-tools.cosmos.esa.int/external/xmm_user_support/documentation/sas_usg/USG/}}.
First, we used the Extended Source Analysis Software package (ESAS)
tasks to unveil the distribution of the extended emission from
NGC\,5189. The ESAS tasks apply restrictive selection criteria of
events, reducing the possible contamination from the astrophysical
background, the soft proton background, and solar wind charge-exchange
reactions, all with important contributions at energies
$<$1.5~keV. The final net exposure times after processing the EPIC
data with the ESAS tasks are 32.2~ks, 41.07~ks, and 44.6~ks for the
pn, MOS1, and MOS2 cameras.

Individual background-subtracted, exposure-corrected EPIC pn, MOS1,
and MOS2 images were created and merged together. We created EPIC
images in the soft (0.3--0.7~keV), medium (0.7--1.2~keV), and hard
(1.2--4.0~keV) bands. The resultant X-ray images of each band as well
as a color-composite image are presented in
Figure~\ref{fig:NGC5189_XRAYS}.

To study the physical properties of the hot gas in NGC\,5189, we have
extracted background-subtracted EPIC spectra.  The data were
reprocessed using the SAS tasks {\it epproc} and {\it emproc} to
produce the EPIC pn and MOS event files.  Lightcurves were created in
the 10--12~keV energy range for each camera and were examined to
search for periods of high background levels.  We rejected time
intervals where the background count rate was higher than
0.2~counts~s$^{-1}$ for both MOS cameras and 0.5~counts~s$^{-1}$ for
the pn camera. The resulting exposure times after this process were
36.64~ks, 53.27~ks, and 55.39~ks for the pn, MOS\,1 and MOS\,2,
respectively, thus reducing the useful observing time by 40-60\%.

\section{Results}

\subsection{Distribution of the hot gas in NGC\,5189}

The {\it XMM-Newton} EPIC images reveal the presence hot gas within
NGC\,5189. Figure~\ref{fig:NGC5189_XRAYS} shows that the
X-ray-emitting gas is mainly distributed toward the E and W lobes with
lower surface brightness in the central region.  The extended emission
is dominated by the soft (0.3--0.7~keV) X-ray band with some
contribution from the medium (0.7--1.2~keV) band. There is no
contribution to the extended emission from the hard (1.2--4.0~keV)
band. We note that the CSPN is marginally detected in the soft X-ray
band. Background or field sources can be seen in the immediate
vicinity of NGC\,5189, in particular, two bright and hard sources lie
just south of the W lobe.
 
To produce a clear distribution of the X-ray-emitting gas we excised
all point sources from the soft X-ray image using the CIAO
\citep[Version 4.9;][]{Fruscione2006} task {\it dmfilth}. The
resulting image is used in the color composite image shown in
Figure~3, along with the H$\alpha$ and [O\,{\sc iii}] {\it HST} images
and the {\it WISE} 12~$\mu$m image. Figure~3 left panel shows that the
hot gas in NGC\,5189 fills the lobes detected by the [O\,{\sc iii}]
emission. Thus, the extent of the hot bubble (shocked, fast wind) does
not reach the N and S knots.  The right panel in Figure~3 shows that
the X-ray emission might be anti-correlated with the emission from the
near-IR as detected by {\it WISE}, suggesting some absorption due to
the presence of dust-rich torus around the CSPN, but the detection of
very soft X-ray emission from the CSPN suggests otherwise (see
discussion section).

Finally, we have overplotted the contours of the extended emission on
the right panel of Figure~3. Faint emission is marginally detected
toward the tips of the E and W lobes.

\begin{figure*}
\begin{center}
  \includegraphics[angle=0,width=\linewidth]{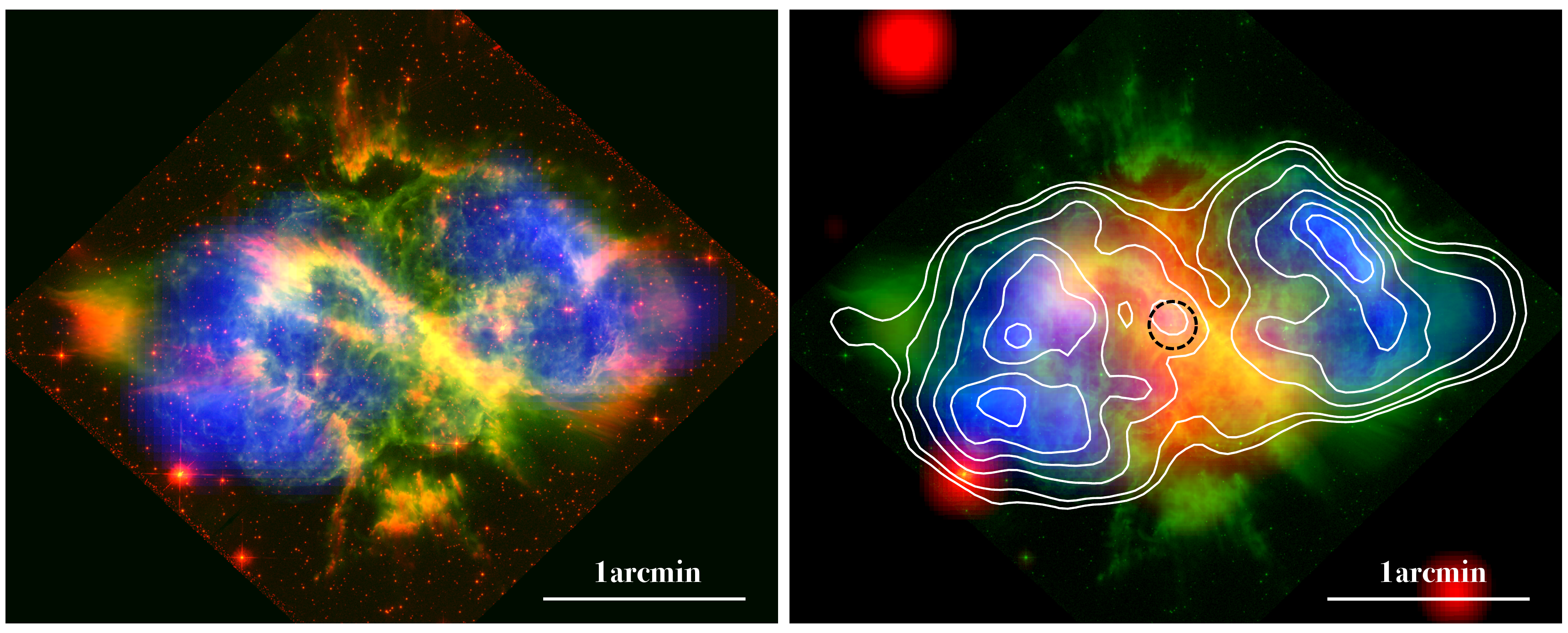}
\label{fig:NGC5189_RGB}
\caption{Color-composite images of NGC5189. Left: red and green
  correspond to the {\it HST} H$\alpha$ and [O\,{\sc iii}],
  respectively. Right: Red and green correspond to the {\it WISE}
  12~$\mu$m emission and {\it HST} [O\,{\sc iii}], respectively. The
  circular dashed-line aperture shows the position of the CSPN. In
  both panels the blue represents the extended soft X-ray
  (0.3--0.7~keV) emission. North is up, east to the left.}
\end{center}
\end{figure*}

\subsection{Physical properties of the hot bubble in NGC\,5189}

The background-subtracted EPIC spectra of the diffuse X-ray emission
are presented in Figure~4. The spectra of NGC\,5189 are very soft with
most of the diffuse X-ray emission detected in the 0.3--1.5~keV energy
range and dominated by emission below 0.6~keV. The peak of emission is
present at $\sim$0.36--0.38~keV which may be due to the C\,{\sc vi}
emission line at 0.37~keV (=33.7~\AA). A secondary peak between
0.5--0.6~keV may be due to the O\,{\sc vii} triplet at 0.58~keV
($\approx$22\,\AA).  Some emission in the spectrum around 0.8--1.0~keV
may be related to the Fe complex and/or Ne\,{\sc xi} lines.  The
corresponding count rates are 35.4$\pm$1~counts~ks$^{-1}$ for the pn
camera and 7.7$\pm$0.60~counts~ks$^{-1}$ for the MOS cameras. These
correspond to total counts of 1300$\pm$50~photons, 410$\pm$33~photons,
and 430$\pm$30~photons for the pn, MOS1, and MOS2 cameras,
respectively.

We modeled the {\it XMM-Newton} EPIC spectra with XSPEC \citep[Version
  12.10.1;][]{Arnaud1996} using an absorbed, optically-thin thermal
plasma model ({\it vapec}).  We initially adopted nebular abundances
from \citet{GR2013} but these abundances did not produce a good fit,
so some elements were allowed to vary to improve the fits. For
comparison and discussion, in Table~1 we list the nebular and CSPN
abundances of NGC\,5189 reported by \citet{GR2013} and
\citet{Keller2014}, respectively.

\begin{figure*}
\begin{center}
  \includegraphics[angle=0,width=0.48\linewidth]{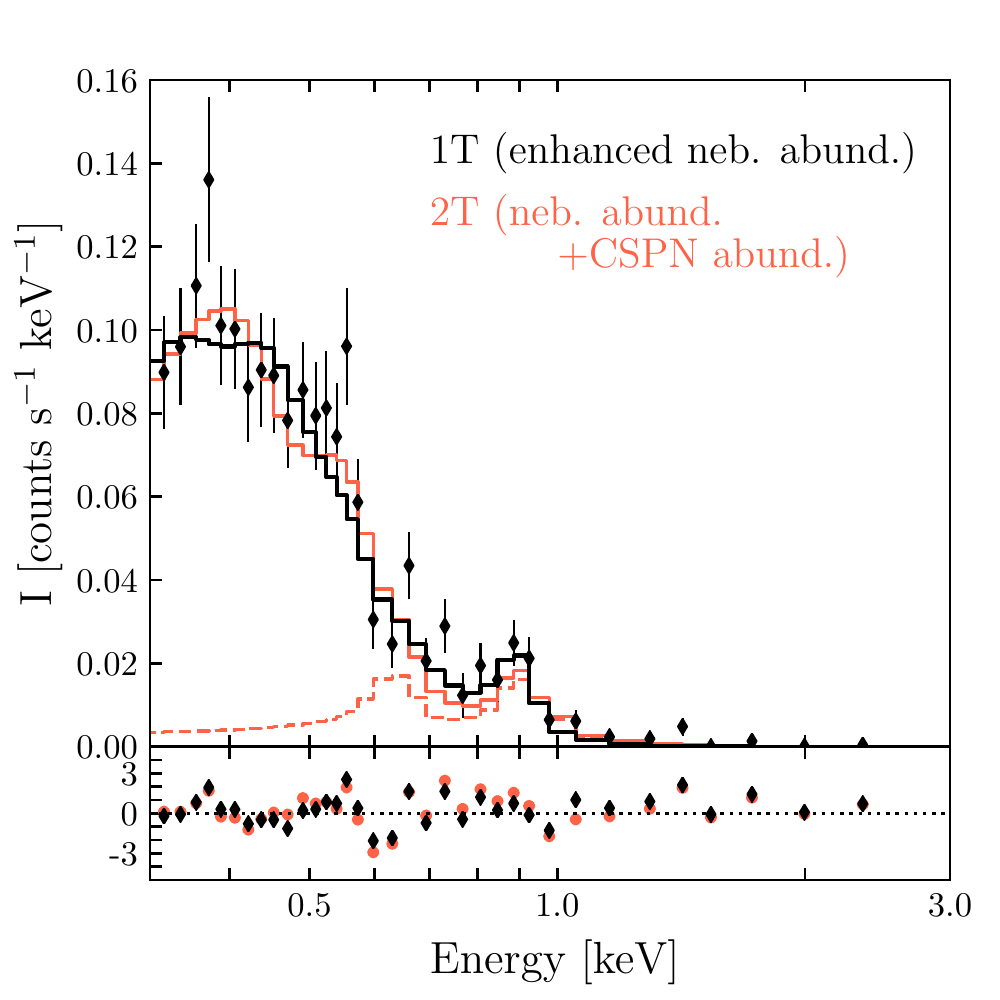}~
  \includegraphics[angle=0,width=0.48\linewidth]{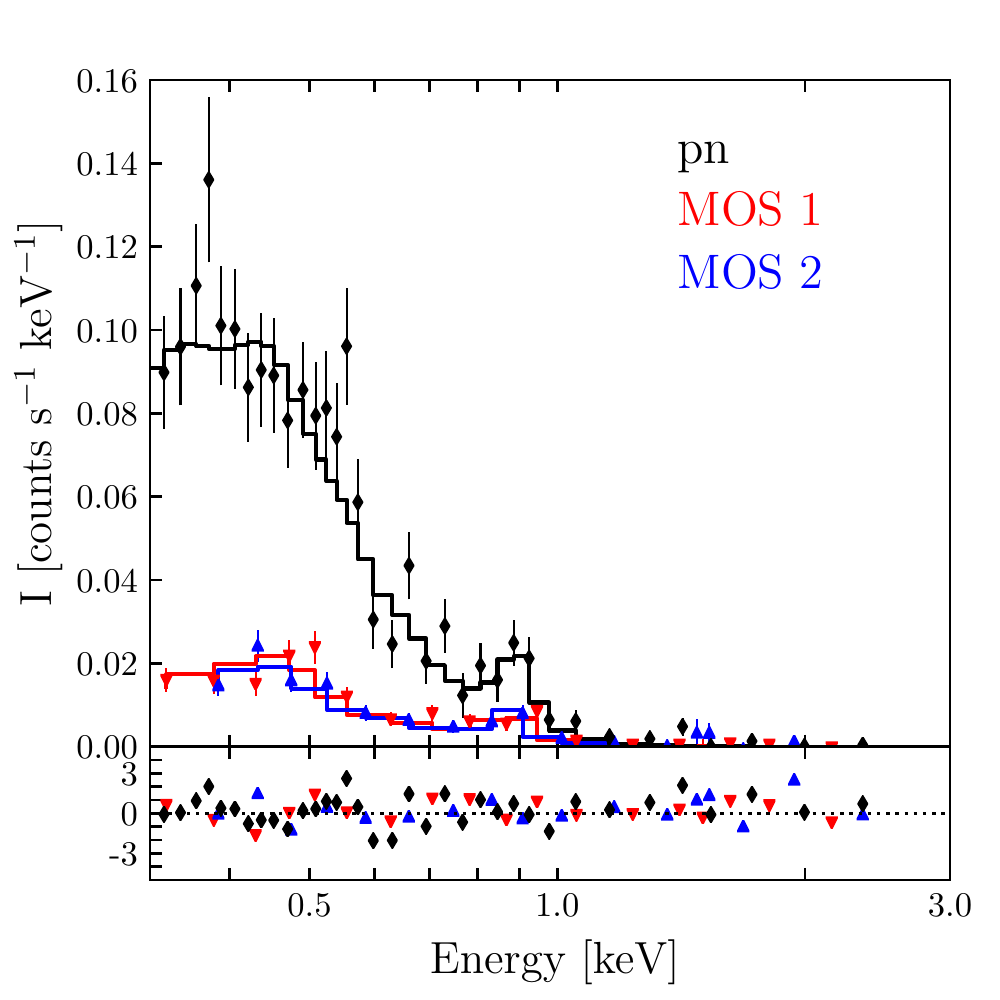}
  \label{fig:spec}
\caption{Background-subtracted EPIC spectra of NGC\,5189. The solid
  lines show the best-fit models to the data while the lower panels
  present the residuals. Left: Best-fit model obtained by fitting a
  one-temperature and a two-temperature plasma models to the EPIC-pn
  spectrum. The dashed line represents the contribution form the
  temperature component with the CSPN abundances. Right: Best-fit
  model obtained by simultaneously fitting a one-temperature {\it
    vapec} plasma model to the three EPIC cameras. }
\end{center}
\end{figure*}

\begin{table}
\centering
\caption{Abundance sets explored for spectral fit of the X-ray emission of NGC\,5189$^\mathrm{a}$}
\label{tab:table1}
\begin{tabular}{cccc}
\hline
Element & NGC\,5189$^\mathrm{b}$     & CSPN$^\mathrm{c}$  & This work$^\mathrm{d}$ \\
\hline
He   &  1.26   & 8.40   & 1.26 \\
C    &  2.90   & 325    & 38$^{+27}_{-15}$ \\
N    &  3.55   & 35     & 3.55  \\
O    &  0.70   & 50     & 0.70  \\
Ne   &  1.50   & 90     & 6.0$^{+3.5}_{-2.4}$\\
S    &  0.80   &        & 0.80  \\
Ar   &  1.40   &        & 1.40  \\
Fe   &  0.001  &        & 0.001 \\
\hline
\end{tabular}
\begin{list}{}{}
\item{$^\mathrm{a}$The abundance values are with respect to solar values \citep{Anders1989}.}
\item{$^\mathrm{b}$From \citet{GR2013}.}
\item{$^\mathrm{c}$From \citet{Keller2014}.}
\item{$^\mathrm{d}$Abundaces obtained for the best-fit model to the diffuse X-ray emission of NGC\,5189.}
\end{list}
\end{table}

The absorption was included with the {\it tbabs} absorption model
\citep[][]{Wilms2000}. A preliminary estimate of the hydrogen column
density of $N_\mathrm{H}=1.9\times10^{21}$~cm$^{-2}$ is obtained from
the $E(B - V)$=0.324~mag \citep[see][]{Danehkar2018,GR2012}, but this
parameter was allowed to vary. Finally, we note that we binned the
spectrum to a minimum of 50~counts per bin.

The first attempt to fit the the EPIC-pn spectrum with nebular
abundances did not yield a good fit ($\chi^{2}>3$). The fit was
improved by leaving the carbon and neon abundances as free
parameters. Other elements were allowed to vary, but they did not
result in any significant improvements to the fit. For example, the
nitrogen abundance converged to values close to its nebular value, so
it was fixed to the nebular values.

The best-fit model ($\chi$=1.36) to the EPIC-pn spectrum resulted in a
plasma temperature of $kT$=(0.14$\pm0.02)$~keV
($T$=1.6$\times$10$^{6}$~K) with an hydrogen column density of
$N_\mathrm{H}$=(2.1$\pm$0.6)$\times$10$^{21}$~cm$^{-2}$, the latter is
consistent with the $E(B-V)$ measurements. The carbon and neon
abundances resulted in 33$^{+35}_{-19}$ and 6.8$^{+4.2}_{-3.4}$ times
their solar values \citep[][]{Anders1989}. The absorbed and intrinsic
fluxes in the 0.3--3.0~keV are
$f_\mathrm{X}=(5.8\pm1.7)\times10^{-14}$~erg~s$^{-1}$~cm$^{-2}$ and
$F_\mathrm{X}=(8.2\pm2.4)\times10^{-13}$~erg~s$^{-1}$~cm$^{-2}$.  At a
distance of 1.68~kpc \citep[][]{Kime2018} this corresponds to an X-ray
luminosity of
$L_\mathrm{X}=(2.8\pm0.8)\times10^{32}$~erg~s$^{-1}$. This model is
shown in the left panel of Figure~4 as a solid black line.

We also fitted the three EPIC cameras simultaneously and the best-fit
model ($\chi^{2}$=1.08) resulted in a plasma temperature of
$kT$=0.14$^{+0.01}_{-0.02}$~keV with carbon and neon abundances
38$^{+27}_{-15}$ and 6$^{+3.5}_{-2.4}$ times their solar values,
respectively. The absorbed and intrinsic fluxes of this model are
$f_\mathrm{X}$=(5.8$\pm$1.7)$\times$10$^{-14}$~erg~s$^{-1}$~cm$^{-2}$
and $F_\mathrm{X}=(8.9\pm2.2)\times$10$^{-13}$~erg~s$^{-1}$~cm$^{-2}$,
with a corresponding luminosity of
$L_\mathrm{X}=(3.0\pm0.7)\times10^{32}$~erg~s$^{-1}$. This model is
shown in the right panel of Figure~4.

In order to assess possible differences between the west and east
lobes of NGC\,5189, we extracted a spectrum from each region and
modeled each separately. Their best-fit models are consistent with
that obtained for the entire X-ray emission. Thus, within error bars,
no spectral variations are detected from the different lobes.

The high C abundance and plasma temperature suggest a carbon-enriched
hot bubble in NGC\,5189. As an additional test, we tried a
two-temperature plasma emission model for the EPIC-pn spectrum. We set
plasma to nebular abundances and another one to CSPN abundances from
Table~1. This model ($\chi^{2}$=1.57) resulted in plasma temperatures
of $kT_\mathrm{neb}$=0.09~keV and $kT_\mathrm{CSPN}$=0.34~keV. This
model is shown in the left panel of Figure~4.  The contribution from
the plasma with abundances as those from the CSPN is also shown as a
dashed line.  We note that this spectral model predicts that the CSPN
should emit primarily from 0.5--1.0~keV, however, there appears to be
little to no emission from the CSPN in this energy range according to
Figure~\ref{fig:NGC5189_XRAYS}.

\subsection{X-ray properties of the CSPN (WD\,1330$-$657)}

The CSPN of NGC\,5189, WD\,1330-657, is marginally detected in the
soft X-ray band (see Fig.~2). We have extracted an EPIC-pn spectrum of
this source. The background has been extracted from a region around
the central source to eliminate any possible contribution from the hot
bubble. The resulting background-subtracted EPIC-pn spectrum is
presented in Figure~5. The spectrum is soft and, similar to the X-ray
spectra of the extended emission, it peaks at energies below
0.5~keV. The net count rate in the 0.3--2.0~keV band is
0.86$\pm$0.21~counts~ks$^{-1}$. This is a total count of
32$\pm$7~photons.

We fit the EPIC-pn spectrum with a one-temperature thermal plasma
model ({\it vapec}) with abundances set to those reported for
WD\,1330$-$657 (Table~1) and with the $N_\mathrm{H}$ set to that
estimated from the $E(B-V)$ measurements.  The best-fit model
($\chi^{2}$=1.20) resulted in a plasma temperature of
$kT=(0.09^{+0.03}_{-0.07})$~keV ($T$=10$^{6}$~K). The observed and
intrinsic fluxes in the 0.3--2.0~keV energy range are
$f_\mathrm{X}=(1.1\pm0.3)\times10^{-14}$~erg~s$^{-1}$~cm$^{-2}$ and
$F_\mathrm{X}=(2.2\pm0.7)\times10^{-14}$~erg~s$^{-1}$~cm$^{-2}$,
respectively. This corresponds to a luminosity of
$L_\mathrm{X,CSPN}=(7.5\pm1.7)\times10^{30}$~erg~s$^{-1}$.

According to the best-fit model to the stellar spectrum of
WD\,1330-657 presented by \citet{Keller2014}, the central star of
NGC\,5189 has a bolometric luminosity of
$L=5010$~L$_{\odot}$\footnote{Were we have rescaled the bolometric
  luminosity obtained by \citet{Keller2014} to the distance of
  1.68~kpc.}. Thus, WD\,1330-657 has a $L_\mathrm{X}/L$ of
$\sim(3.7\pm0.8)\times10^{-7}$. The X-ray temperature and luminosity
suggest that the X-ray emission from the vicinity of the CSPN is
mainly due to shocks in the stellar winds from WD\,1330-657 and not a
spun-up binary companion.

\begin{figure}
\begin{center}
  \includegraphics[angle=0,width=\linewidth]{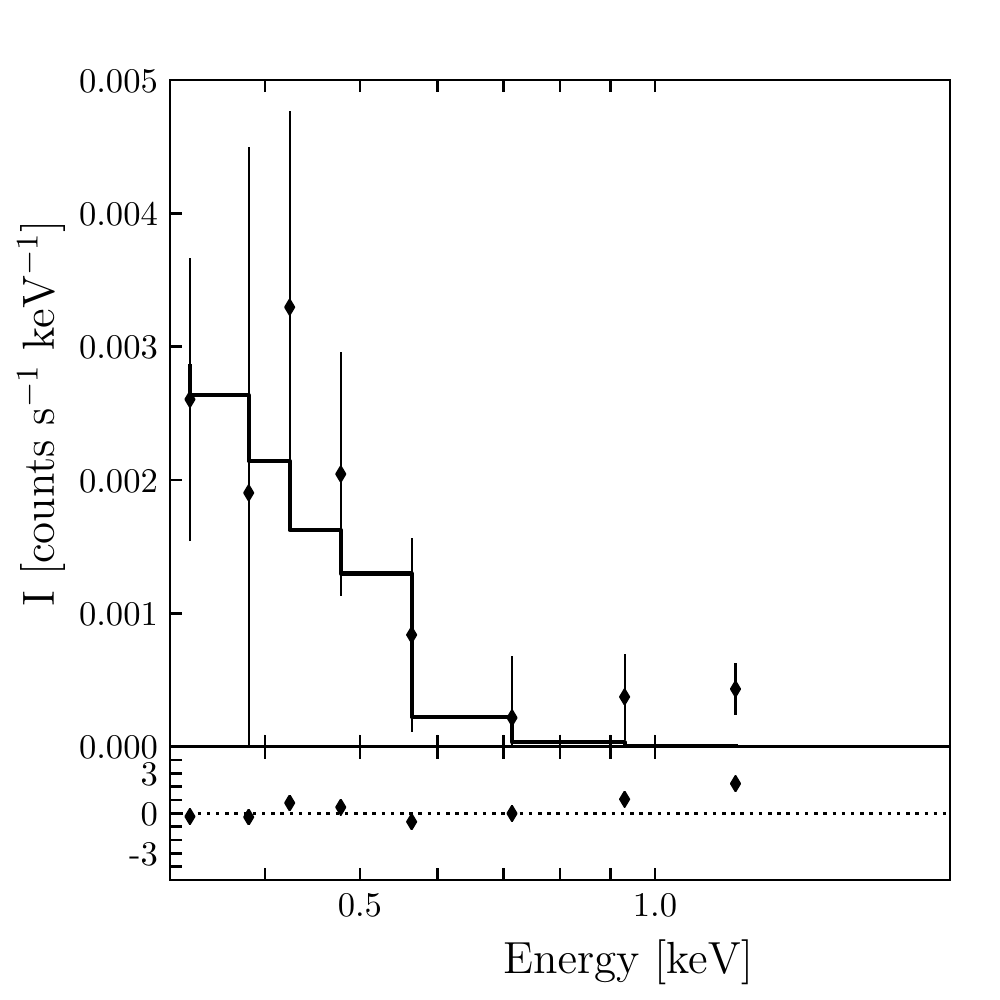}
\label{fig:spec_CSPN}
\caption{EPIC-pn background-subtracted spectra of the CSPN of
  NGC\,5189, WD\,1330-657.}
\end{center}
\end{figure}

\section{Discussion}

The discovery of extended, soft X-ray emission in NGC\,5189 reveals
true extension of the effects of the fast wind from the [WR]-type star
WD\,1330-657.  The X-ray-emitting material is primarily detected
inside the lobes of NGC\,5189 as defined by the [O\,{\sc iii}]
emission.  The N and S knots, which reside beyond the [O\,{\sc iii}]
shell, are not interacting with the hot bubble emission powered by the
fast CSPN stellar wind as suggested by \citep{Sabin2012}.  Some soft
(0.3--0.7 keV) X-ray emission is detected from the edges of the E and
W lobes and might suggest a possible leakage of the hot bubble beyond
the [O\,{\sc iii}] boundary, but we note that \citet{Sabin2012} find
no clear evidence of kinematic signature of fast components such as
jets.

The X-ray-emitting material appears limb-brightened towards the outer
edges of the E and W lobes. In other words, the regions close to the
star have a lower emissivity. The lower emissivity could be intrinsic
to the source or caused by overlying absorption.  Indeed,
\citet{Sabin2012} have suggested that the IR emission detected around
the CSPN of NGC\,5189 might be due to the presence of a dust-rich
structure. However, an inspection of the {\it ISO} spectrum of
NGC\,5189 reveals only a modest contribution from dust amid bright
emission lines at 10.5~$\mu$m and 15.5~$\mu$m, likely corresponding to
[S\,{\sc iv}] and [Ne\,{\sc iii}] lines, and which dominate the {\it
  WISE} 12~$\mu$m band. A strong emission line at $\sim26~\mu$m,
possibly due to the [O\,{\sc iv}] or [Fe\,{\sc ii}], dominates the
{\it WISE} 22~$\mu$m band suggesting that the IR emission detected
around the CSPN is likely dominated by highly-ionized emission lines
which might be evidence of evaporation of material around the CSPN
(see following subsection).  This notion is further supported by the
fact that the CSPN is hot
\citep[$T_\mathrm{eff}=165$~kK;][]{Keller2014} and is detected in the
absorption-sensitive soft X-ray band with little evidence for enhanced
absorption from the X-ray spectral fitting.

We suggest that the limb-brightened morphology of the diffuse X-ray
emission in NGC\,5189 might arise from the scenario proposed by
\citet{Akashi2008}. These authors presented numerical simulations in
which a PN is shaped by a bipolar fast wind (a jet) creating a hot
bubble capable of producing soft X-ray emission. Their models predict
that the X-ray emissivity will peak at the outer edges of the lobes
\citep[see figure~1 in][]{Akashi2008}.

\subsection{The origin of the X-ray emission in NGC\,5189}

One of the main results of the {\sc ChanPlaNS} project is that diffuse
X-ray emission from a hot bubble is mainly detected in compact PNe
with radii $<$ 0.2~pc.  The X-ray emitting E and W lobes of NGC\,5189
extend to up to 1.42$^{\prime}$ which, at a distance of 1.68~kpc,
correspond to a hot bubble cavity with a radius of 0.7~pc.  Other
large and highly-evolved PNe in the {\sc ChanPlaNS} survey of similar
size to NGC\,5189 (A\,33, LoTr5, HFG\,1, DS\,1), do not indicate any
diffuse X-ray emission.  Hence, NGC\,5189 is the largest PN with
diffuse X-ray emission and its detection is unexpected.

Although the X-ray emission from evolved PNe are expected to be
enriched with nebular material due to hydrodynamical mixing or the
thermal conduction effect (see Section~1), our spectral analysis has
unveiled that the X-ray-emitting gas in NGC\,5189 is strongly
carbon-enriched. This carbon abundance (38 times its solar value) is
larger than the carbon abundance reported for the nebula (see
Table~1). There are three other cases in which this same scenario is
presented. The young and compact PN BD$+$30$^{\circ}$3639
\citep[][]{Yu2009,Kastner2000} and the so-called {\it born-again} PNe
A\,30 and A\,78 \citep{Guerrero2012,Toala2015}, which also harbor
[WR]-type carbon-rich CSPN. Since BD$+$30$^{\circ}$3639 is young, it
has been suggested that its X-ray emission is mainly due to the
shocked, unmixed stellar wind \citep[e.g.,][]{Yu2009,Toala2016}.

In the born-again PNe A\,30 and A\,78 \citep[and possibly also in
  NGC\,40; see][]{Toala2019}, it has been suggested that the X-ray
emission originated as a result of a relatively recent
\citep[$\sim$1000~yr;][]{Fang2014} {\it very late thermal pulse}, in
which hydrogen-deficient, carbon-rich material was ejected into their
evolved PN \citep[$\sim$11,000~yr;
  see][]{Herwig1999,MillerBertolami2006}. In A\,30, A\,78 and
NGC\,5189, the X-ray emission is dominated by the C\,{\sc vi}
Ly$\alpha$ emission line at 0.37~keV (33.7~\AA) and their CSPNe share
similar luminosities and effective temperatueres
\citep[log$_{10}$($L$/L$_{\odot}$)=3.7--3.8,
  $T_\mathrm{eff}$=120--170~kK;
  e.g.,][]{Guerrero2012,Toala2015,Keller2014}, which are too luminous
for their old and evolved PNe. These properties suggest that NGC\,5189
could have experienced the born-again evolutionary path like that of A
30 and A 78.

Although the carbon and neon abundances determined from the extended
X-ray emission from NGC\,5189 are high compared to the nebular
abundances, they are not as high as those of the CSPN (Table~1). This
may be an indicative of the mixing process between the carbon-rich hot
bubble and the nebular material, which causes the dilution of the
highly carbon enriched stellar material abundance of the shocked
X-ray-emitting material.

Alternatively, a binary interaction (similar to that of a nova
eruption on an O-Ne-Mg WD) has also been invoked in order to explain
the specific C/O abundance ratio from born-again PNe
\citep[][]{Wesson2003,Wesson2008}. \citet{Lau2011} summarized the
abundance determination from born-again PNe and from different models
in their Table~1. We note that the mass fractions of helium, carbon,
oxygen and neon from the CSPN of NGC\,5189 are not consistent with
those reported for nova predictions, but are more consistent with the
late thermal pulse scenario.

WD\,1330-657 hosts a companion with an orbital period of 4.04~d
\citep[][]{Manick2015}, although this companion might have been
involved in the early shaping of NGC\,5189
\citep[e.g.,][]{Manick2015,Frank2018,Chamandy2018}, it may not be
involved in powering the extended X-ray emission nor the emission from
the vicinity of the CSPN. The relatively soft nature of the tentative
X-ray detection of the CSPN (WD\,1330-657) and its estimated
log$_{10}(L_\mathrm{X}/L)\sim-7$ ratio, suggest that self-shocking
winds might be the dominant factor in the production of X-rays from
the CSPN as opposed to a spun-up companion
\citep[]{Montez2010,Montez2015}.  The spectrum and physical properties
($T_{\rm X}$, $L_{\rm X}$, $L_{\rm X}/L_{\rm bol}$) of the CSPN are
similar to those of the CSPN of PN\,K\,1-16, which was detected
serendipitously by both {\it Chandra} and {\it XMM-Newton}
\citep{Montez2013}.  For PN\,K\,1-16, where only the CSPN is detected,
the X-ray emission is consistent with a self-shocking wind in a
carbon-rich environment.

\section{Conclusions}

We presented the discovery of extended X-ray emission from the PN
NGC\,5189 around the [WR]-type star WD\,1330-657.  NGC\,5189 harbors a
carbon-enriched X-ray-emitting plasma with the largest hot bubble
radius of all PNe detected thus far in X-rays. Our findings can be
summarized as follows:

\begin{itemize}

\item The distribution of the X-ray-emitting gas delimited by the
  [O\,{\sc iii}] emission (Figure~3). The extent of the hot bubble
  does not reach the N and S knots suggesting that there is no
  dynamical interaction between the knots and the shell with the N and
  S knots. Hence, the cometary shape of these knots are likely due to
  the photoevaporation by UV flux.

\item Analysis of the spatial distribution of the X-ray-emitting gas
  in NGC\,5189 suggests the presence of possible hot bubble blowouts
  at the farthest regions of the W and E lobes.

\item The spectrum of the diffuse X-ray-emitting hot bubble gas in
  NGC\,5189 is soft with the bulk of the X-ray emission detected at
  energies below 0.6~keV. This is dominated by the C\,{\sc vi}
  emission line at 0.37~keV (33.7~\AA).

\item The EPIC spectra could not be modelled with nebular nor CSPN
  abundances. We found that the X-ray-emitting material has carbon and
  neon abundances $\sim$38 and 6 times solar values, respectively.

\item The presence of the carbon-rich hot bubble in this extended and
  old PNe suggests an extreme physical process such as that of the
  born-again PNe A\,30 and A\,78. In this scenario, these PNe are
  thought to undergo a very late thermal pulse that further ejects
  carbon-rich material found in the hot bubble. This scenario is
  supported in NGC\,5189 by the fact that the estimated abundances of
  the hot bubble reside between those of the PN and the CSPN.

\item Although the binary companion of the CSPN might have been
  involved in the shaping of NGC\,5189, it does not seen to be
  involved in the marginally detected X-ray emission from
  WD\,1330-657. The origin of its X-ray emission is similar to that of
  K\,1-16 and other soft X-ray emitting CSPN, and likely caused by
  self-shocking winds such as those observed from hot massive stars.

\end{itemize}

To summarize, we suggest that NGC\,5189 was initially shaped by a
common-envelope process, but soon after the CSPN might have
experienced a very late thermal pulse that resulted in the WR
characteristics of WD\,1330-657 and which subsequently powered the
formation of the large hot bubble. Hence, the extended X-ray emission
is the result of the mixing between the carbon-rich hot bubble and the
nebular material from NGC\,5189.

Follow-up {\it XMM-Newton} observations, with a longer exposure and
higher signal-to-noise, are needed to better determine the detailed
spatial distribution and physical characteristics of the extended
X-ray emission. Future observations with higher spectral and spatial
resolution, such as those that will be provided by $Athena+$, will
help accurately determine the abundance of the X-ray-emitting gas in
NGC\,5189 and its CSPN. Improved C and Ne abundances will further
probe the carbon enrichment origins of NGC\,5189 proposed
here. Furthermore, spectroscopic and high-resolution IR observations
of the central region of NGC\,5189 would help assess the presence of a
dusty toroidal structure around WD\,1330$-$657 and its potential role
in shaping the nebular structure and absorbing the X-ray emission.

\acknowledgements
JAT acknowledges support from the UNAM DGAPA PAPIIT
project IA100318. RMJ and MK acknowledge support from the {\it
  Chandra} X-ray Center (CXC), which is operated by the Smithsonian
Astrophysical Observatory (SAO) for and on behalf of NASA under
contract NAS8-03060.  The authors thank M.A. Guerrero for fruitful
discussion that enhanced the presentation of the present work.

\facilities{{\it XMM-Newton}.}

\software{SAS \citep{Gabriel2004}, XSPEC \citep{Arnaud1996}, CIAO \citep{Fruscione2006}.}



\end{document}